\pdfoutput=1
\documentclass[11pt,twoside]{article}
\usepackage{asp2010}
\resetcounters

\begin{document}

\title{(Sn)DICE: A Calibration System Designed for Wide Field Imagers}
\author{N.~Regnault$^1$, E. Barrelet $^1$, A.~Guyonnet$^1$, C.~Juramy$^1$, P.-F.~Rocci$^1$, L.~Le~Guillou$^1$, K.~Schahman\`eche$^1$, F. Villa$^1$}
\affil{$^1$ Laboratoire de Physique Nucl\'eaire et de Hautes-Energies, Barre~12-22~1$^{\mathit{er}}$~\'etage, 4 place Jussieu, 75252 Paris CEDEX 05.}
% \affil{$^2$Institution Full Address for Author2}
% \affil{$^3$Institution Full Address for Author3}}

\begin{abstract}
  Dark Energy studies with type Ia supernovae set very tight
  constraints on the photometric calibration of the imagers used to
  detect the supernovae and follow up their flux variations. Among the
  key challenges is the measurement of the shape and normalization of
  the instrumental throughput. The DICE system was developed by
  members of the Supernova Legacy Survey (SNLS), building upon the
  lessons learnt working with the MegaCam imager.  It consists in a
  very stable light source, placed in the telescope enclosure, and
  generating compact, conical beams, yielding an almost flat
  illumination of the imager focal plane.  The calibration light is
  generated by narrow spectrum LEDs selected to cover the entire
  wavelength range of the imager. It is monitored in real time using
  control photodiodes.  A first DICE demonstrator, SnDICE has been
  installed at CFHT. A second generation instrument (SkyDICE) has been
  installed in the enclosure of the SkyMapper telescope. We present
  the main goals of the project.  We discuss the main difficulties
  encountered when trying to calibrate a wide field imager, such as
  MegaCam (or SkyMapper) using such a calibrated light source.
\end{abstract}

\section{Introduction}

Modern precision cosmology, such as the measurement of the Dark Energy
equation of state with type Ia supernovae (SNe~Ia) \cite[see e.g.][and
references therein]{sullivan_2011, conley_2011, guy_2010,
  kessler_2009} sets very tight constraints on the accuracy of the
flux calibration of the imagers.  Indeed, the measurement of the
luminosity distance of high-redshift (resp. low-redshift) supernovae
is primarily performed with the redder (resp. bluer) bands of the
imagers. Hence, measuring cosmological parameters with SNe~Ia
ultimately boils down to comparing fluxes measured either with red and
blue passbands and it is fundamental to control the intercalibration of
the imager passbands. Taking full advantage of statistics and quality
of SN~Ia measurements requires to control this intercalibration with
an accuracy of a fraction of a percent.

The current photometric calibration techniques rely on observations of
spectrophotometric stellar calibrators.  Establishing such primary
standards is notoriously difficult, as one has to anchor astronomical
observations to a physical flux scale. One of the best efforts so far
is the work of the CALSPEC team \cite[][and references
therein]{CALSPEC}.  The CALSPEC flux scale relies on NLTE {\em models}
of three (and now five) pure hydrogen white dwarfs.  These stars are
the primary standards used to calibrate the flux response of the HST
instruments, in particular STIS and NICMOS, which are used in turn to
extend the CALSPEC library by adding secondary spectrophotometric
standards. The SNLS and SDSS-II surveys have chosen to anchor their
flux calibration on this so-called HST white dwarf flux scale
\citep{betoule_2012, regnault_2009, holtzman_2008}. The uncertainty
that affects the spectrum of the primary standards is however
difficult to assess. As the precision of the calibration efforts
improves, it seems increasingly important to check these stellar
calibrators using laboratory standards.

The wavelength positioning of the survey passbands has also a sizeable
impact on the cosmological parameter measurements, as shown for example
in table 9 of \cite{conley_2011}.  The required wavelength accuracy on
the filter cut-offs is as low as a fraction of a nanometer. Passband
models are usually derived from pre-installation test-bench
measurements of the imager optical components. This is not entirely
satisfactory, as filters may evolve over time time as shown for
example in \cite{doi_2010}, and it seems necessary to be able to
measure and follow up {\em in situ} the instrument passbands.

With these requirements in mind, several groups have sought to develop
instrumental calibration systems, i.e.  light sources that can
illuminate the telescope pupil with well characterized light. By
todays standards, ``well characterized'' means that the calibration
beam has been mapped using Si photodiodes procured from an institute
of standards such as the american Institute of Standards and
Technology (NIST).  Many different designs have been proposed over the
last few years, and quite a few are now being tested on various wide
field imagers.  In what follows, we describe the DICE system. DICE
stands for Direct Illumination Calibration Experiment. It consists in
a very stable point-like source, generating conical beams that deliver
a quasi-uniform illumination on the focal plane. Two such systems have
been built so far. A first prototype was installed at the Canada
France Hawaii Telescope (CFHT) in order to calibrate the 1 deg$^2$
MegaCam imager. A second generation demonstrator was recently
installed in the enclosure of the 5.7 deg$^2$ SkyMapper imager. We
discuss below the main design (\S \ref{sec:design_considerations}) and
implementation (\S \ref{sec:dice_system}) aspects of the project. We
then describe the test bench procedures that permit to characterize
the light source (\S \ref{sec:test_bench}). Finally, we discuss a few
key problems that arise in the data analysis (\S \ref{sec:analysis}).

\section{Design Considerations}
\label{sec:design_considerations}

\subsection{Calibration beam}

\begin{figure}
\plottwo{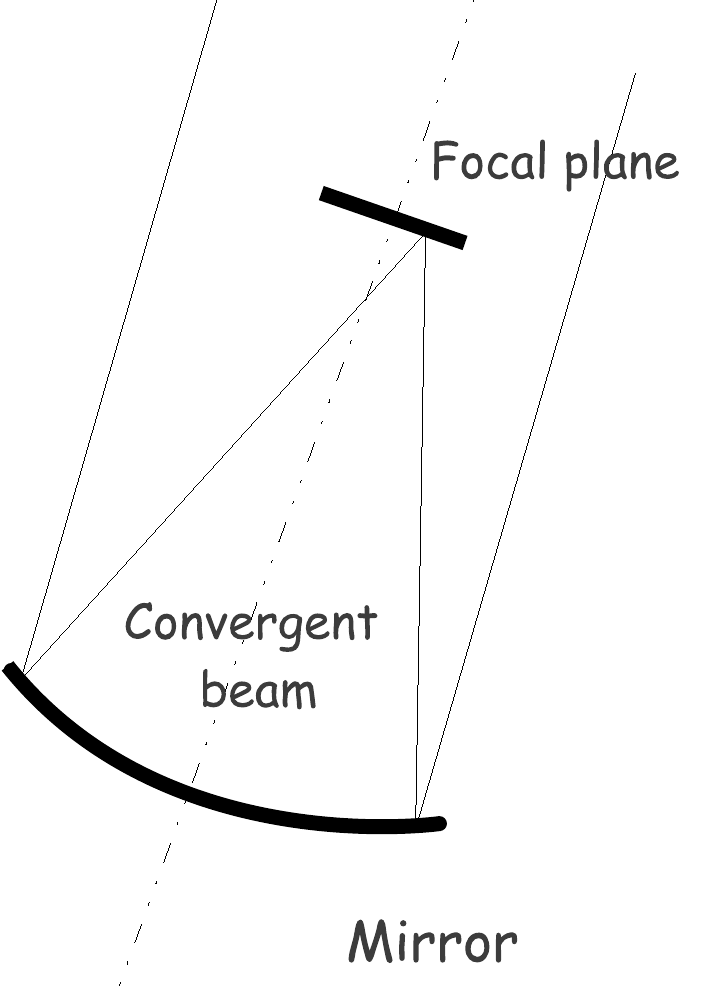}{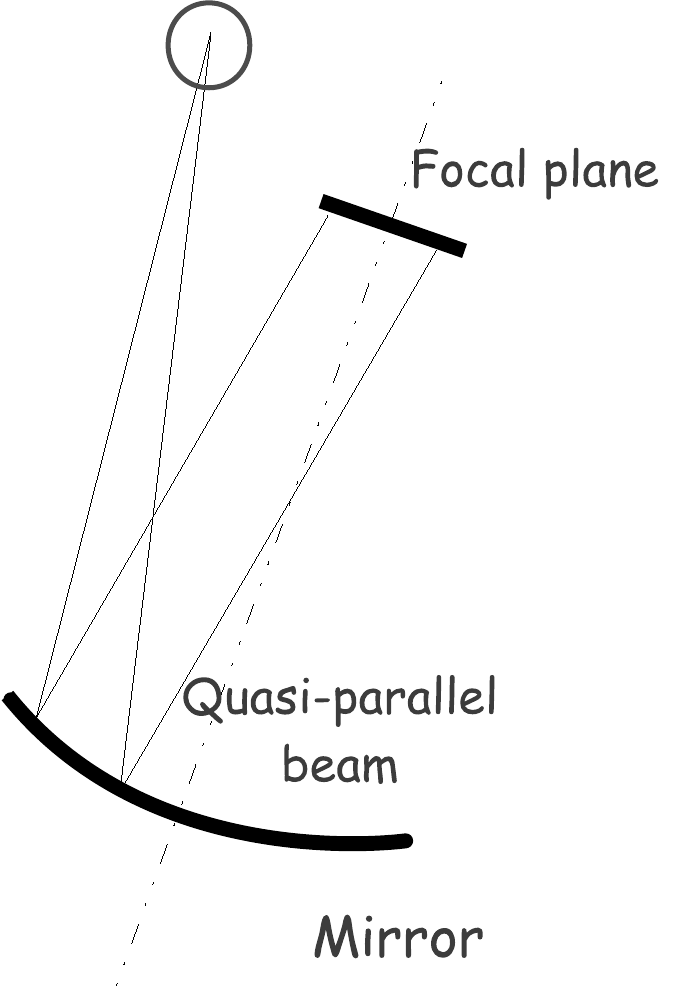}
\caption{Left: telescope illumination by a point source (``science
  beam''). Right: SnDICE calibration beam.\label{fig:calibration_beam}
}
\end{figure}

Ideally, a calibration device should mimic as much as possible the
science objects under study. Since a supernova survey is dealing
primarily with point sources (supernovae and field stars), we should
try and generate quasi-parallel beams, covering the entirety of the
primary mirror (see figure \ref{fig:calibration_beam}). Such a beam
would result in a spot on the focal plane, and we could use the
photometry code in production in the survey photometry pipeline to
estimate its flux, thereby avoiding the systematic errors that arise
from using different flux estimators.

Unfortunately, building a good artificial star turns out to be
difficult. For SnDICE, we deliberately opted for a different design,
(figure \ref{fig:calibration_beam}). SnDICE is a point source, located
in the dome, a few meters away from the telescope primary mirror,
close to the object plane. The source generates a conical,
quasi-lambertian beam, of aperture $\sim 2^{\mathrm o}$, slightly
larger than the telescope angular acceptance. Such an illumination
results in an almost uniform focal plane illumination.

As shown on figure \ref{fig:calibration_beam}, the calibration beam is
radically different from the science beam. In particular, the angular
distribution of the light rays that hit the various optical surfaces
(e.g. the interference filters) is not comparable. However, this
specific calibration beam has at least one very nice property: it is much
simpler than the science beam, in the sense that each pixel sees
photons that came through a unique path. In other terms, there is a
one-to-one relationship between the focal plane elementary surface
elements and the calibration beam elementary solid angles. As we will
see, such a property makes it very simple to predict the focal plane
illumination, once one knows the beam radian intensity map.

\subsection{Light emitters}

Another important design element is the choice of narrow-spectrum
light emitting diodes (LED) as light emitters. LEDs are known to be
extremely stable, as long as they are fed with stable currents. It it
is relatively easy today to build current sources stable at a few
$10^{-5}$ over a temperature range of a few degrees. Hence, with some
care, it is relatively easy to build a light source that can deliver
very stable beams over long durations.  The diversity of
narrow-spectrum LEDs available on the market permits to cover the
entire spectral range of silicon imagers, from the near UV to the near
infrared (see figure \ref{fig:led_spectral_coverage}). 

LED do not emit monochromatic light. The width of typical LED spectra
is of about $\delta \lambda / \lambda \sim 5 - 7 \%$. Hence, as
always, there is a trade-off, as we chose to sacrifice wavelength
precision in favor of high-quality and high-stability illumination.
This makes sense, since what one actually needs is a {\em follow-up}
more than a {\em measurement} of the filter cutoff positions. The
filter transmissions are indeed well measured prior to installation.

LEDs come with just one caveat: their emission properties vary with
temperature.  As temperature increases, the LED emission efficiency
drops by up to 0.5\% / $^{\mathrm o}$ C, and the mean wavelength of the
emitted light shifts redwards by about 0.1 \AA / $^{\mathrm o}$ C. As
will be discussed in \S \ref{sec:test_bench}, these variations are
generally linear and always extremely reproducible. As a consequence, once
each emitter has been well characterized, one only needs to implement
a real time follow-up of the source temperature to account for these
effects.

\section{The DICE System}
\label{sec:dice_system}

The SnDICE light source is a $130 \mathrm{mm} \times 130 \mathrm{mm}
\times 300 \mathrm{mm}$ box, pierced with 25 asymmetric holes. It
contains 24 calibration LEDs mounted on a radiator, located on the
back of the device, approximately 260 mm from the front face. The
light beams exit through circular apertures of diameter 9-mm. The
beams are carefully shaped using a series of masks designed to kill
most of the stray light.  Off-axis control photodiodes, located close
to the front face, monitor in real time the light delivered by each
LED. The mechanical design of SkyDICE, the demonstrator installed in
the SkyMapper enclosure, is almost identical, except that the LED head
is shorter, in order to generate wider (3 degree) conical beams.

\begin{figure}
\plotone{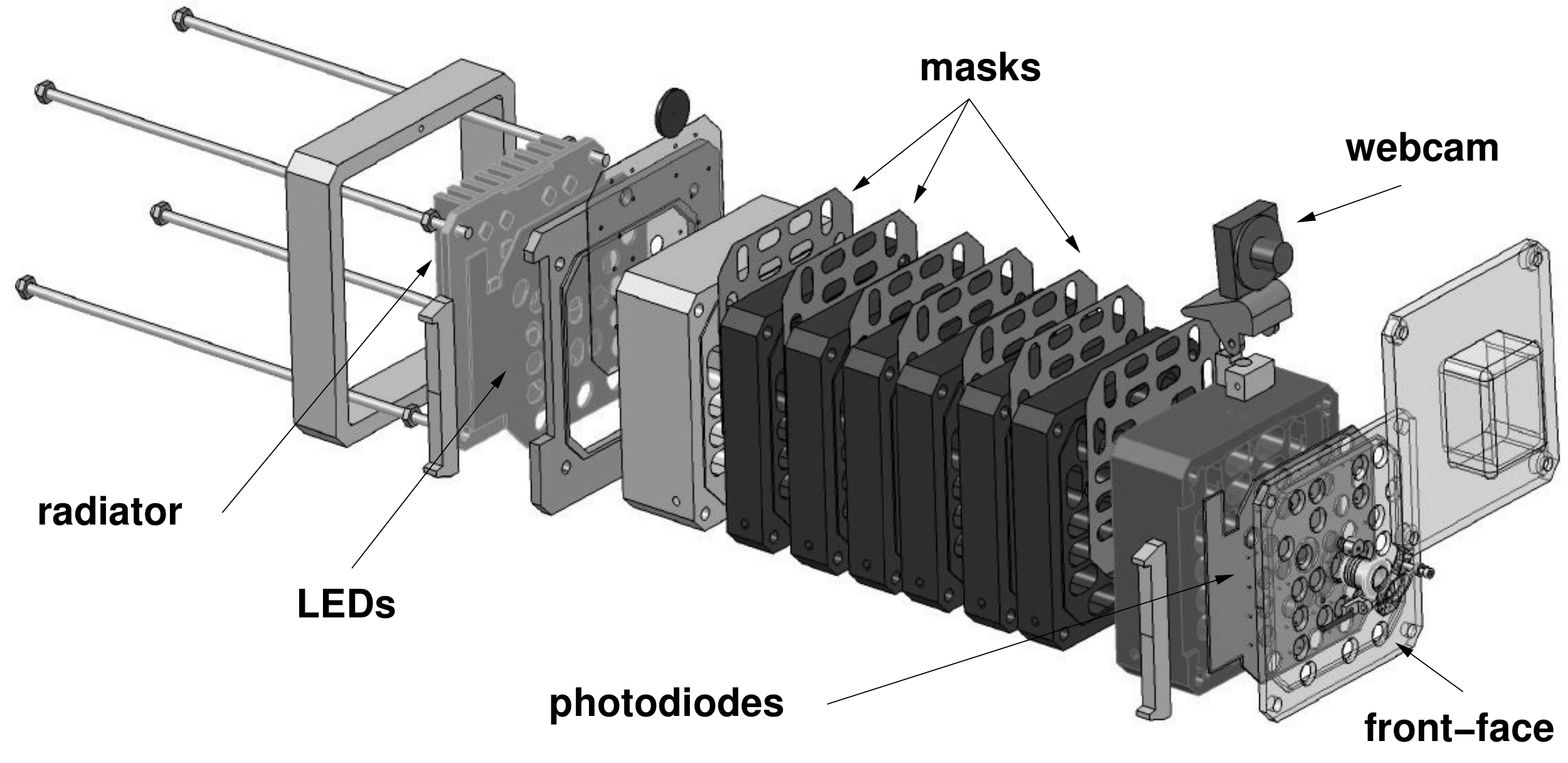}
\caption{Sketch of the DICE light source. It consists in a modular
  box, made of 8 (almost) identical aluminum pieces. pierced with
  holes to let the light through. The LEDs are mounted on a radiator,
  on the back of the device. Close to the front face, control
  photodiodes (one for each calibration channel) monitor the light
  delivered by the LEDs.}
\end{figure}

\begin{figure}
\plottwo{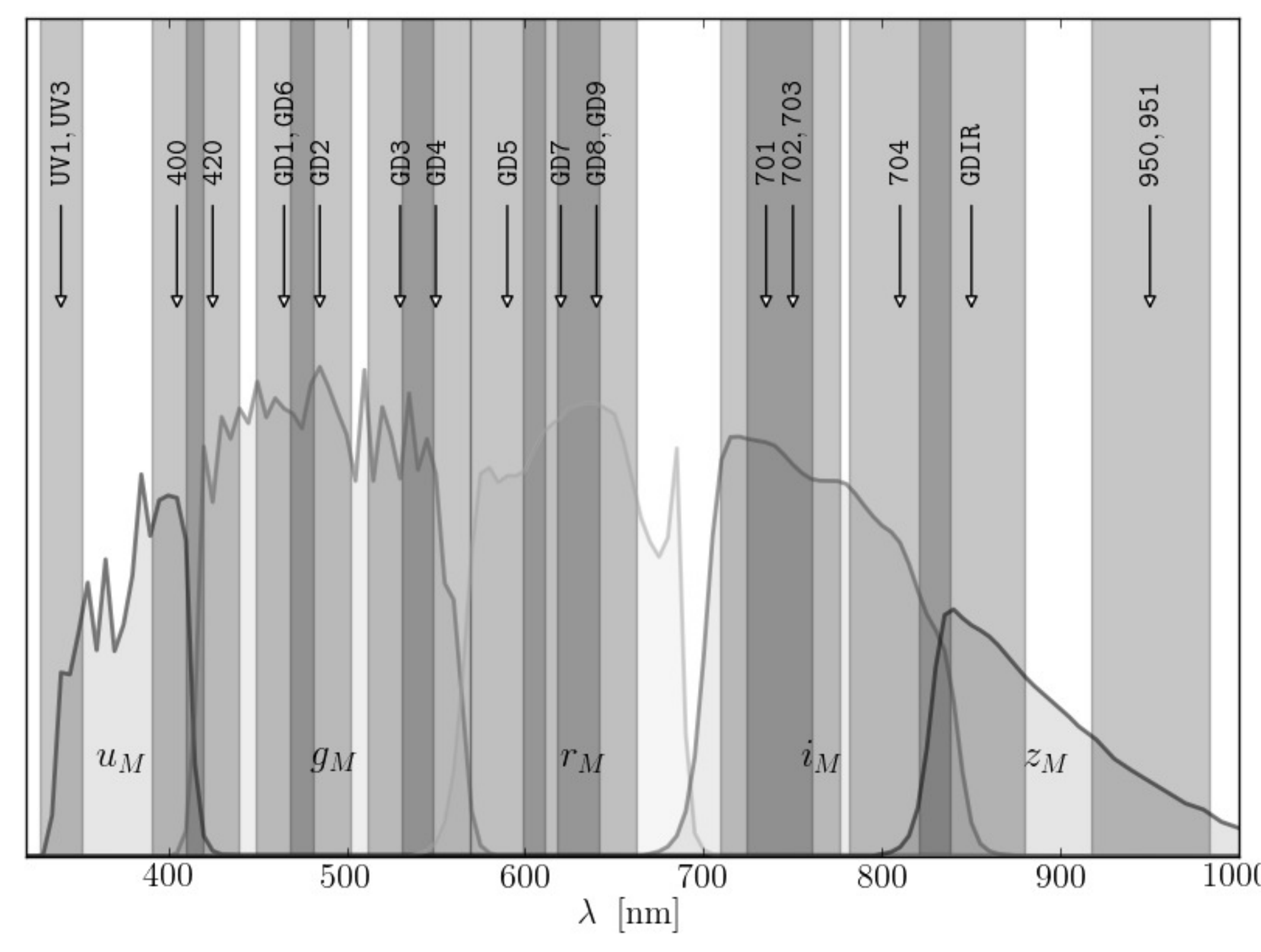}{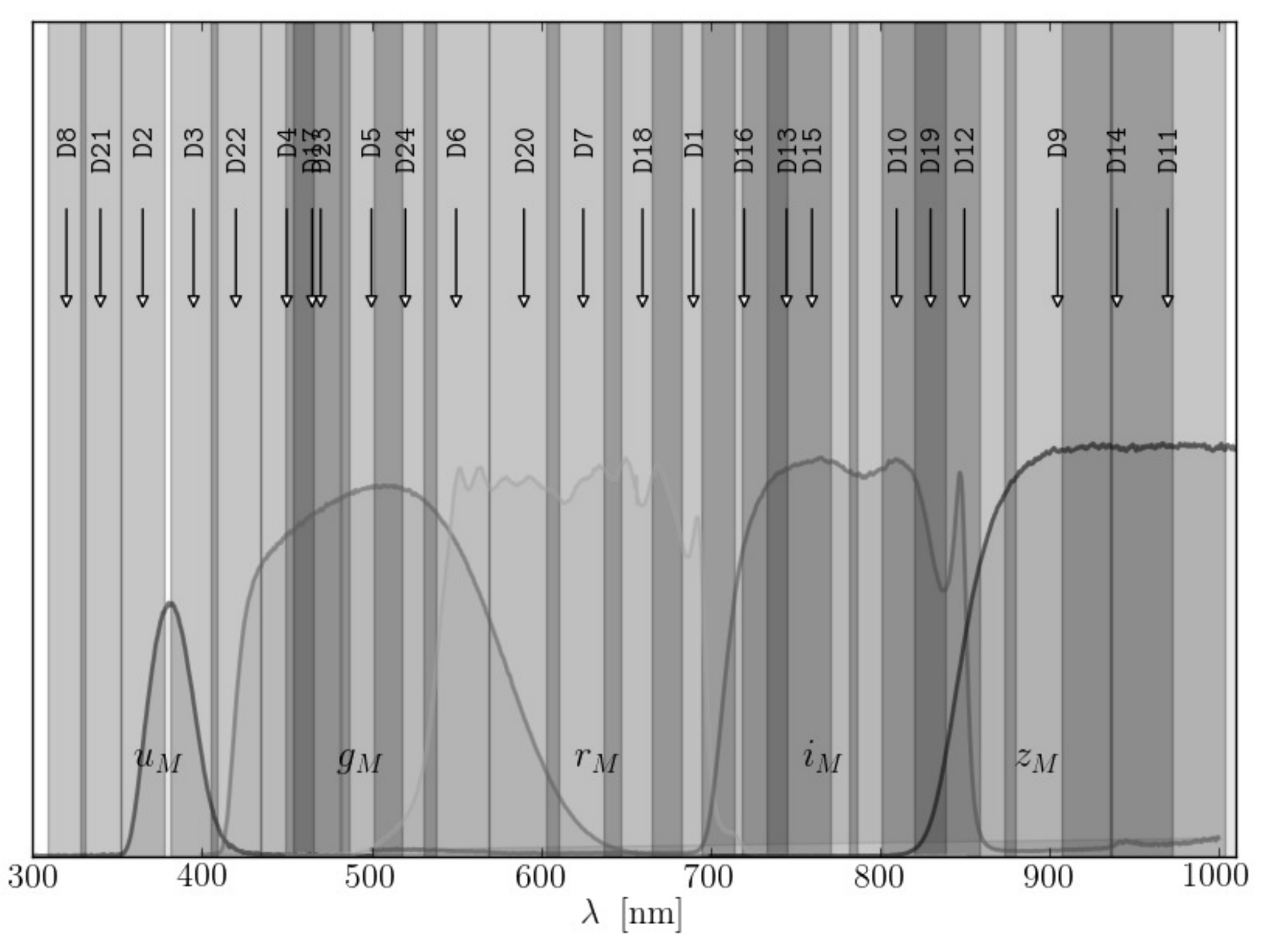}
\caption{Left: SnDICE spectral coverage of the MegaCam passbands. The
  arrow indicate the LED central wavelengths $\bar{\lambda}$. The shaded region
  delimitates the region $ 0.95 \bar{\lambda}< \lambda < 1.05 \bar{\lambda}$ to
  give an idea of the LED spectrum extension.  Right: SkyDICE coverage
  of the SkyMapper passbands.\label{fig:led_spectral_coverage} }
\end{figure}

Figure \ref{fig:led_spectral_coverage} shows the quality of the
spectral coverage obtained from SnDICE (left) and SkyDICE (right). The LEDs have been
chosen so that each filter is sampled by at least three LEDs: one
close to the maximum, and two covering the filter cutoffs. As can be
seen, the diversity of LEDs available in 2007, at the time SnDICE was
designed, does allow to sample well the red cutoff of the MegaCam $r$
filter, as well as the blue cutoff of the $i$-filter. Also, few LEDs
were available in the infrared. Two years later, the diversity of LEDs
had dramatically improved, and the spectral coverage provided by
SkyDICE is excellent.

The source can be oriented using an altitude and an azimuth motor. By
moving simultaneously the LED head and the telescope, one is able to
scan various locations of the primary mirror, keeping both instruments
aligned. A 25th LED called the {\em artificial planet} is used to
control the relative alignment of SnDICE with the telescope. This
channel is equipped with a convergent lens that reshapes the light
into a (quasi) pencil-beam. Planet exposures display a bright ($\sim
100$ pixel wide) spot, along with a series of {\em ghosts} generated
by reflexions within the telescope optics. The position of the main
spot is a direct measurement of the relative orientation of the planet
beam and the telescope optical axis. The relative positions of the
ghosts encode information on (1) the distance between the planet beam
and the telescope optical axis and (2) the alignment of the telescope
optics --the WFC lenses, in particular.

The DICE system is designed to deliver very stable beams over long
durations. The LEDs are operated at very low currents and should not
be subject to significant evolution. Also, there are no intermediate
optical surfaces between the emitters and the telescope primary mirror
--besides the LED encapsulant. This limits the effect of optical
surface ageing. Finally, a set of redundant controls has been
implemented: the LED currents and temperatures are monitored in real
time, the LED flux is also measured by control photodiodes. Finally,
cooled, large area control photodiodes, designed to sense the typical
low focal plane illuminations, of a fraction of nW / cm$^{-2}$, are
placed on the telescope, and provide and additional, independant
measurement of the calibration flux delivered by the source.

During a typical data taking session, the dome aperture tracking
system is disconnected and the telescope points inside the dome
towards the source. The relative alignment and positions of both
instruments are controlled with a series of alignment exposures taken
with the {\em planet} beam. Then, sequences of calibration exposures
are taken.  The source and the telescope are then moved, in order to
cover a different mirror area, and the same calibration sequence is
repeated. Some calibration sequences are designed to test the
repeatability of the readout electronics. They consists in up to 50
exposures taken with the same LED, at regular intervals. The source
itself being very stable --and monitored-- this allows one to study
potential variations of the imager gains over time. Other calibration
sequences consists in series of exposures taken with different LEDs,
with and without filter. Such a dataset permits to monitor the
telescope transmissions -- filter normalization and filter cutoffs.

\section{Test Bench Studies}

Prior to installation, the light source must be characterized on a
spectrophotometric test bench. The goal is (1) to measure the spectral
energy distribution of each LED, as a fonction of temperature and (2)
to map the beam radiant intensity (in W / sr) -- also at various
temperatures. All measurements are performed for several values of the
LED currents, although, in operations, we try not to vary the LED
currents, and adapt the total illumination delivered to the focal
plane by varying the exposure time. The nominal current was adapted
for each LED, in order to yield a few thousand photoelectrons per
pixel and per second on the focal plane.

The test bench is placed in a $2 \mathrm{m} \times 2 \mathrm{m} \times
3.5 \mathrm{m}$ dark enclosure. The enclosure walls are insulated, and
its temperature can be regulated from 0$^{\mathrm{o}}$C (typical
temperature on site) to 25$^{\mathrm{o}}$ C. The bench is not strictly
speaking thermalized, as significant temperature gradients can still
be measured after a few hours of operations. However, the temperature
of all bench elements is monitored using PT1000 thermistors.

The flux measurements are all performed with a calibrated photodiode,
purchased from an institute of standards. A very common choice is the
1~cm$^2$ Hamamatsu~2281, calibrated at the National Institute of
Standards and Technology (NIST). Following the recommandations of
NIST, it is operated at ambiant temperature, in photoeletric mode (not
polarized in reverse) and read out using a Keithley 6514 feedback
picoammeter. The fluxes measured by the photodiode are of a few
nanowatts, yielding photodiode currents of a few nanoamperes (the
fluxes recorded by the imager focal plane are about 50 times lower).

\begin{figure}
\plotthree{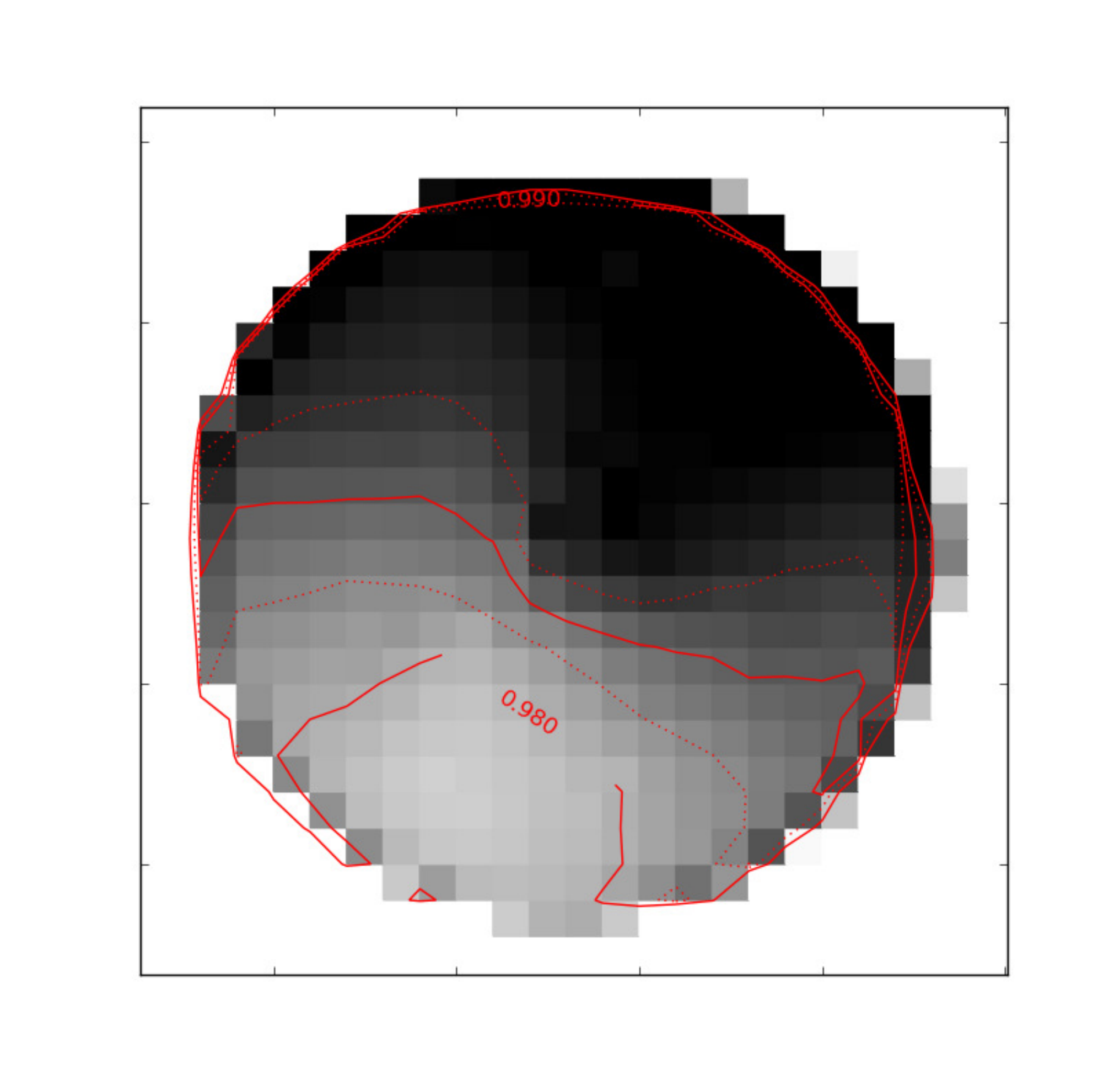}{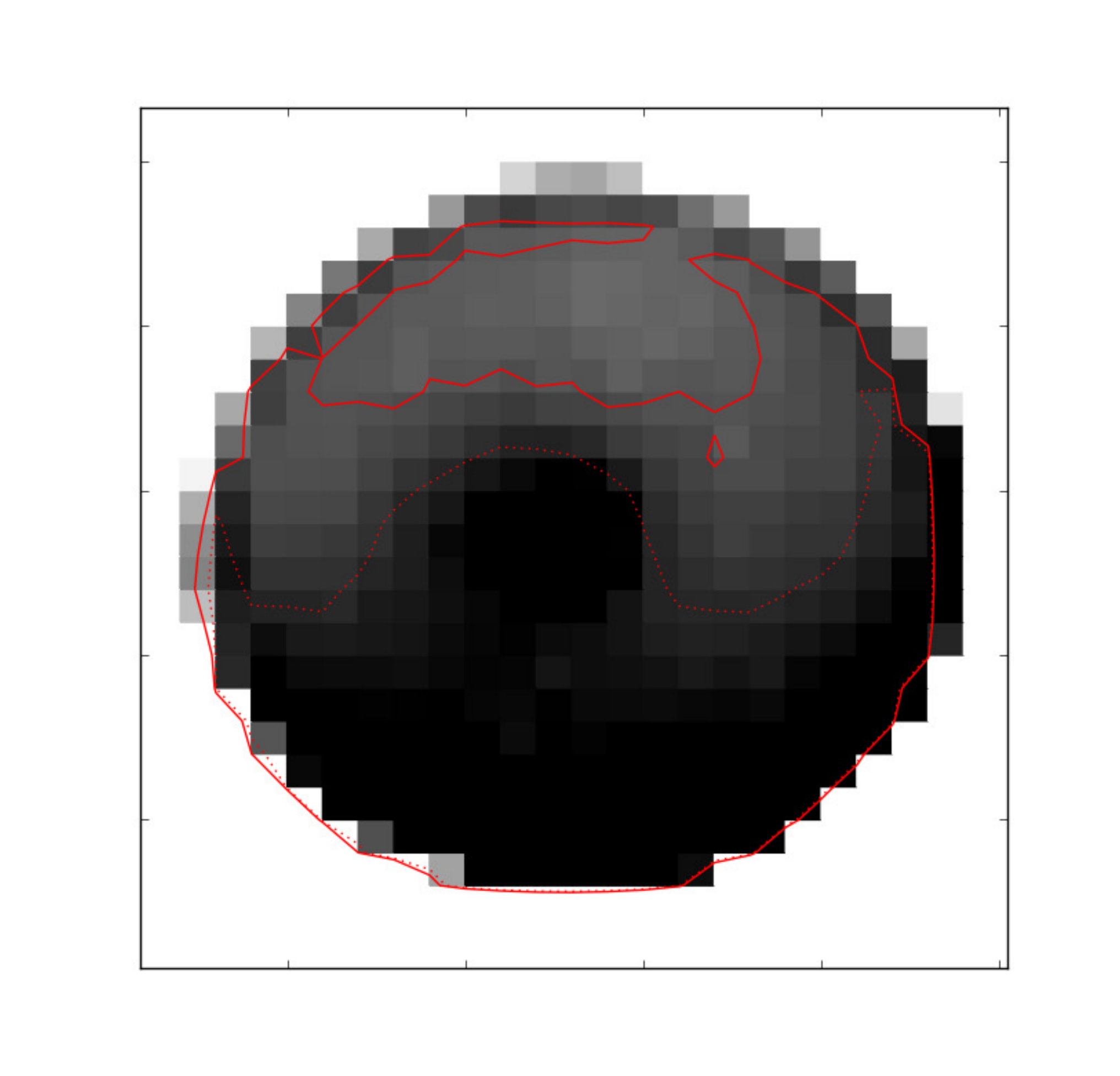}{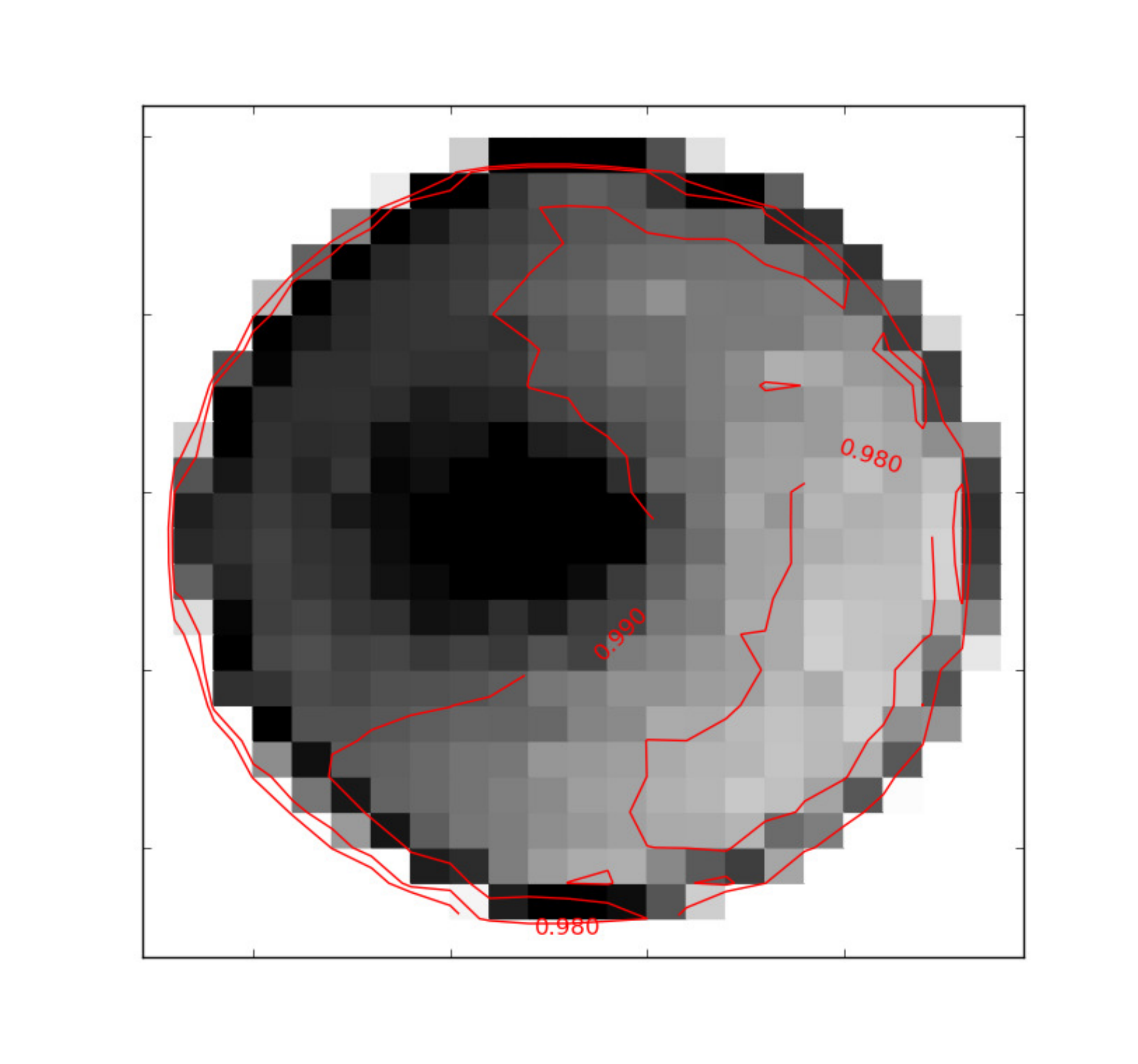}
\caption{Map of the radiant intensity delivered by three different
  LEDs, in the blue, green and near UV. The diameter of each beam is
  of 2 degrees. The non-uniformities shown on the map are of about 1\%
  (peak-to-trough)}
\label{fig:beam_radiant_intensity}
\end{figure}

The bench operations are fully automated. The LED head and the
calibrated photodiode are both mounted on computer controlled linear
tables.  By moving the photodiode in a plane orthogonal to the head
$Z-$axis one can map the beam radiant intensities. Figure
\ref{fig:beam_radiant_intensity} shows three different maps. The beams
are not perfectly lambertian, but display small non-uniformities at
the level of about 1\%. The origin of those non-uniformities, in
particular of the central ``bump'' is unknown. However, they are
extremely well measured, with a relative precision of a few 0.01\% and
do not seem to vary.

\begin{figure}
\plotone{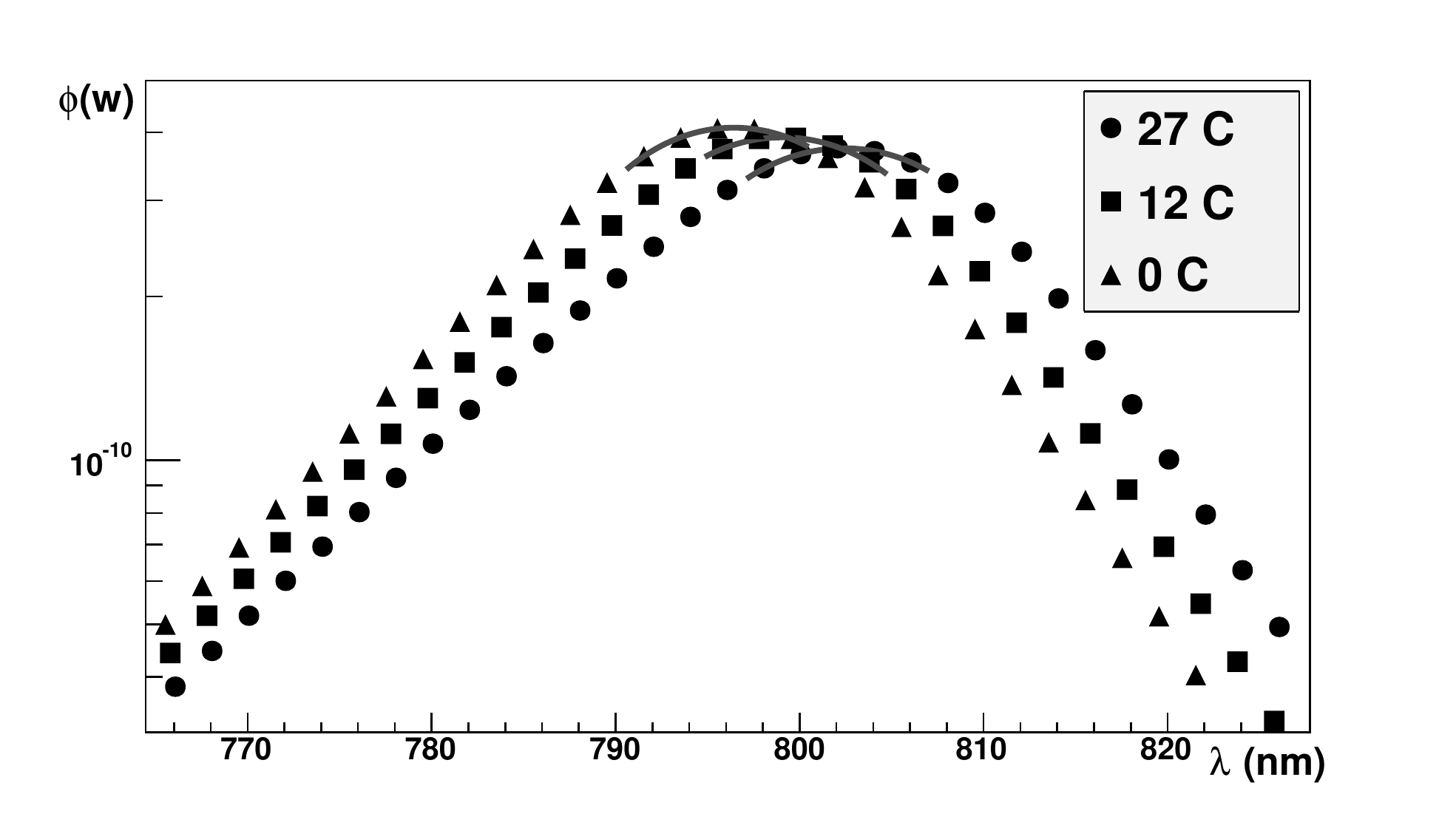}
\caption{Three spectra of the same LED (810 nm), taken at three different temperatures.}
\label{fig:led_spectra}
\end{figure}

By placing a monochromator (a Digikr\"om DK240) between the source and
the calibrated photodiode one can measure the spectrum of each LED. To
do this, the wavelength calibration of the monochromator has to be
performed several times over the full temperature range, using Cd and
Hg lamps, as small but significant variations of the instrument
response with temperature have been found. Taking this into account,
the total wavelength uncertainty is of about 1\AA.  Figure
\ref{fig:led_spectra} shows a typical LED spectrum at three different
temperatures. The behavior described in \S
\ref{sec:design_considerations} is clearly apparent: as temperature
decreases, the flux increases and the spectrum is blueshifted. Also,
the shapes of the spectra seem to change slightly with temperature.

\label{sec:test_bench}

At the end of the process, we have a full spectrophotometric
characterization of the light source. The spectroscopic measurements
give the shape of the spectra; the photometric measurements allow one
to derive their normalization, as a function of the direction with
respect to the soure $Z-$axis.  To derive the LED emission properties
at any given temperature, we interpolate through the measurements. The
fluxes are anchored on the physical flux scale carried by the
photodiode. According to NIST, the uncertainties that affect the
measurement of the photodiode efficiency are of about 0.2\% in the
visible ($400 \mathrm{nm} < \lambda < 950 \mathrm{nm}$). However, NIST
does not yet report the correlations between measurements performed at
different wavelengths. As a consequence, one cannot compute the
uncertainty affecting the relative normalization of LED of different
wavelengths without some amount of guessing.

\section{Analysis of the Calibration Frames}
\label{sec:analysis}

\subsection{Gains \& Imager Follow-up}

The simplest study that can be done with a DICE system, is to detect
and follow-up the short term variations of the imager response -- the
gains in particular. This has been done using long series of exposures
taken with the same LED, taking advantage of the stability of the
light source. Several such studies have been performed with the
MegaCam imager.  They have shown that most of the 72 readout channels
display a stability of about 0.1\% (rms), some 5 channels being much
more unstable, with a variability of 0.5\% to 1\%. 

DICE can not only study the gain short term temporal variations of the
imager response, but also determine with one single exposure, the
step-like variations of the imager response, due to the
amplifier-to-amplifier gain differences, as well as the CCD-to-CCD
quantum efficiency variations. Indeed, the beam radiant intensity
being continuous, the focal plane illumination is also
continuous. Hence, sharp CCD-to-CCD or amplifier-to-amplifier
variations displayed by the calibration frames permit to
intercalibrate the CCD and amplifier responses. Hence, performing DICE
observations on a regular basis allows to capture the temporal and
spatial variations of the imager response (at least the non continuous
part of it).

\subsection{Flux Measurements}

One can expect that the illumination $\phi(\vec{x})$ recorded at a
given location $\vec{x}$ on the focal plane when the telescope is
illuminated by SnDICE is simply the beam radiant intensity map,
multiplied by the response fonction of the telescope:
\begin{equation}
  \phi(\vec{x}) = {\cal B}(\vec{u}) \times T(\vec{x}) \times \left|\frac{\partial \vec{x}}{\partial \vec{u}}\right|
  \label{eqn:illu_0}
\end{equation}
where $T(\vec{x})$ is the transmission of the telescope, at the LED mean
wavelength (i.e. what we are trying to measure), ${\cal B}(\vec{u})$
is the radiant intensity emitted in direction $\vec{u}$ by the source
(measured on a test bench). Knowing the telescope optics, one can
easily compute the geometrical factor $\left|\frac{\partial
    \vec{u}}{\partial \vec{x}}\right|$, and derive $T(\vec{x})$. 

Unfortunately, the situation is slightly more complicated, as the
focal plane is also hit by stray light, most of it coming from
internal reflections within the telescope wide-field corrector (WFC).
The contamination level varies from a few percents to a few per-mil as
a function of the focal plane position. Its level depends on the
transmissions and reflectivities of the surfaces encountered on the
telescope optical path, and also on the curvatures of the various
optical surfaces. Hence, the expected flux is rather a sum on all possible 
paths $p$ that light may follow (yielding non-negligible fluxes):
\begin{equation}
  \phi(\vec{x}) = \sum_p {\cal B}(\vec{u}) \times T_p(\vec{x}) \times \left|\frac{\partial \vec{x}}{\partial \vec{u}}\right|_p
  \label{eqn:illu}
\end{equation}
where $T_p(\vec{x})$ is the product of the reflectivies and transmissions
encountered by light over its path through the telescope optics. The
term $\left|\frac{\partial \vec{x}}{\partial \vec{u}}\right|_p$
encodes the geometrical response of the optics, {\em for the path $p$
  under consideration}. As for direct light, it can be computed using
a model of the telescope optics. 

We have built a model of the telescope optics, using information
provided by the telescope teams. The model itself consists in a
raytracer that predicts the impact position on the focal plane, as a
function of the incident ray position and orientation, and this, for
any chosen light path (i.e. not limited to the direct light). In
addition, an empirical focal plane model was built, using the WCS
information contained in the science image headers, in order to
predict the impact positions in pixels (rather than in
millimeters). Finally, the last component of the model --surprisingly
the most difficult one to build-- is a series of geometrical functions
that can predict the relative positions and orientations of
the source and the telescope with a sufficient accuracy, for any given exposure.

\begin{figure}
\plotone{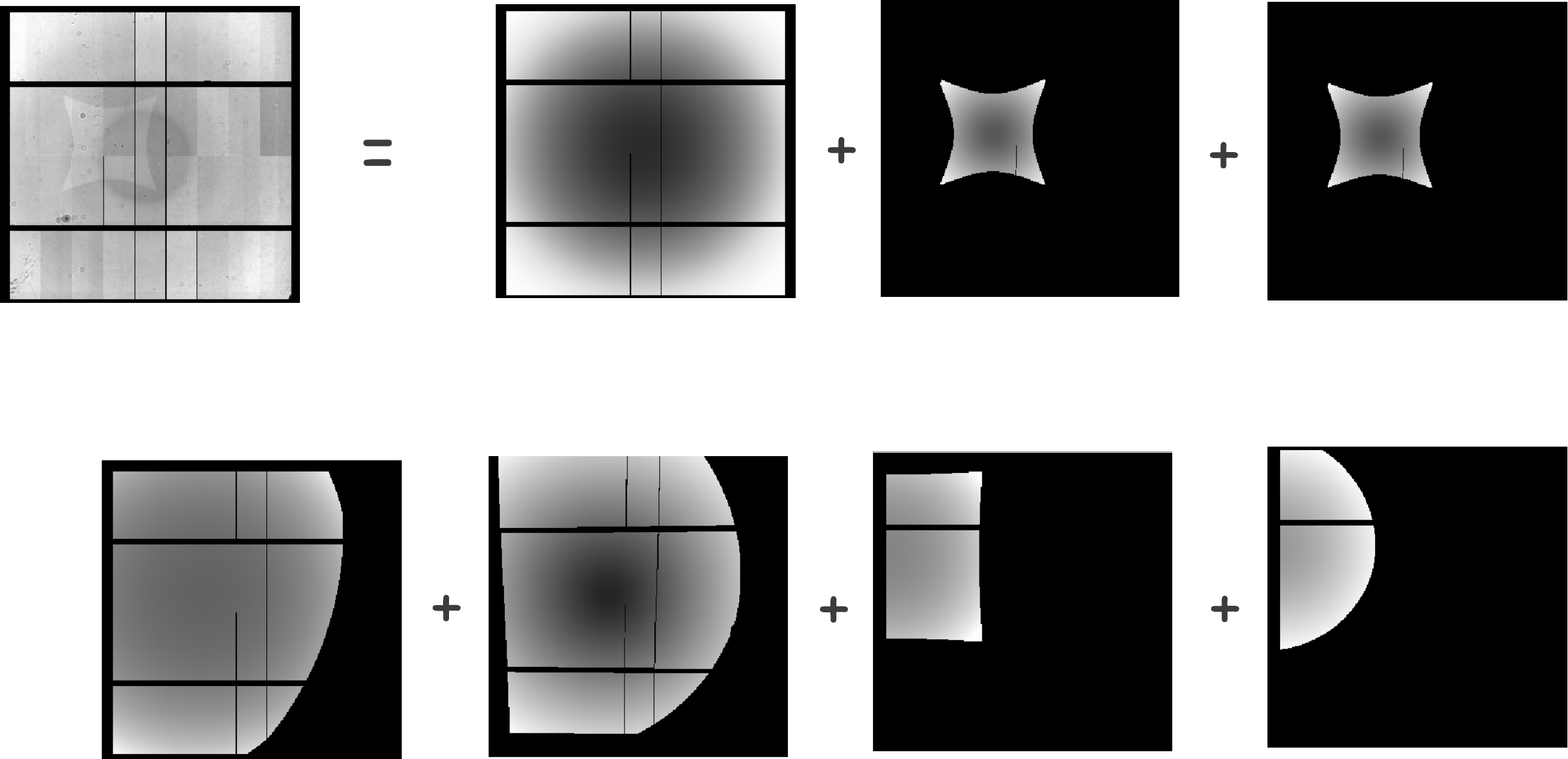}
\caption{Decomposition of a calibration frame (upper left) into the
  sum of (1) the direct light (2) ghosts due to the L4, L3 and L2
  lenses of the wide field corrector.  Note that it is not a
  Monte-Carlo: we do not propagate photons, but we rather use
  equations \ref{eqn:illu_0} and \ref{eqn:illu} to directly compute
  the focal plane illumination.}
\label{fig:sum_of_ghosts}
\end{figure}

These predictions were tested and tuned on the {\em planet} exposures,
where the ghosts and main spot are well separated.  From this, it is
relatively easy to compute the expected illumination, for any given
ghost. Figure \ref{fig:sum_of_ghosts} gives an idea on how one can
decompose a calibration frame as the sum of a series of contributions,
including the direct light and the main ghosts. From this, it is then
possible to estimate the contamination of each calibration frame, and
to reconstruct the telescope transmission.

Note that the problem discussed above, in particular the challenges
posed by contamination by ghosts are, at first order, independant of
the light source design.  If we were using an extended illumination
device, such as a screen, predicting the focal plane illumination
would probably be more difficult: first one would have to measure (and
follow up) the screen {\em radiance} (in W / sr / m$^2$); second, to
predict the focal plane illumination, one would have to compute an
integral over the entire screen, which is computationally much more
demanding.

\subsection{Applications}

Both DICE projects have allowed us to gather very rich calibration
datasets whose analysis is still underway.  Detailed studies of the
imager stability were performed. Preliminary determinations of the
MegaCam passbands were obtained. Over the course of this analysis, we
have developed accurate models of the instruments (telescope + image).
These models were tuned on the {\em planet} exposures, which gave
additional information on the telescope optics (lens positions and
reflectivities).

Also, several applications of DICE --not anticipated initially-- are
currently under study. In particular, it seems that, since DICE
permits to estimate the gains of the imager, and to extract the
instrument response to the direct light (i.e. to separate the direct
light from the ghosts), we can build a ghost-free flat field from a
series of DICE exposures.  This would be a very useful application.
Indeed, for most wide field imagers, the contamination of twilight
flats by stray light alters very significantly the uniformity of the
instrument response, as discussed for example in
\cite{betoule_2012,regnault_2009}.

\section{Conclusion}

Instrumental calibration is today a very active subject. Many
different light source projects have been proposed and it is quite
difficult to predict which design is going to prevail in future
surveys. DICE is one of those attempts. It consists in a compact,
versatile, inexpensive LED based calibrated light source, that can be
placed in the dome of virtually any telescope.  As a point like source
placed at a finite distance, it generates a conical beam that yields a
quasi-uniform focal plane illumination. This is complemented by a
pencil beam that allows (1) to control the relative positions and
orientations of the source and the telescope and (2) to study in
detail the ghost contamination.  The stability of the light source,
permits to carry out a daily monitoring of the telescope and imager
response.  The simplicity of the calibration beams, combined with the
complementarity pencil beam / conical beam allows to derive an
accurate estimate of the telescope throughput, taking into account the
contamination by stray light.

This project does not address the estimation of the atmospheric
transmission, which is an important (and in some bands, highly
variable) component of the telescope effective throughput. Many
contributions to these proceedings are dealing with this subject.

\acknowledgements The authors would like to thank the CFHT Corporation
and the SkyMapper team for their support. The CFHT staff, in
particular Greg Barrick, Jean-Charles Cuillandre, Kevin Ho, Derek
Salmon and Jim Thomas provided tremendous amount of help during the
installation of the device and the data taking runs that followed. On
the SkyMapper side, Brian Schmidt, Peter Conroy, Patrick Tisserand,
Annino Vaccarella, and Colin Vest provided invaluable support at all
stages of the SkyDICE project. Finally, we acknowledge support from
the LPNHE engineering department, notably P. Bailly, J. Coridian,
C. Goffin, H. Lebbolo, A. Guimard, P. Repain, A. Vallereau and 
D. Vincent.

\bibliography{regnault_n}

\end{document}